\documentclass[twocolumn]{openjournal}
\usepackage[T1]{fontenc}
\usepackage{xcolor}
\usepackage{textgreek}
\usepackage[english]{babel}

\usepackage{hyperref}
\hypersetup{
    unicode, 
    colorlinks=true,
    linkcolor=linkcolor,
    citecolor=linkcolor,
    filecolor=linkcolor,
    urlcolor=linkcolor,
}
\definecolor{linkcolor}{rgb}{0.0,0.3,0.5}
\usepackage{tensind}
\tensordelimiter{?}
\DeclareGraphicsExtensions{.bmp,.png,.jpg,.pdf}
\usepackage{verbatim}
\usepackage[normalem]{ulem}
\usepackage{orcidlink}
\usepackage{soul}
\usepackage{graphicx}
\usepackage{amsmath}
\usepackage{bm}
\usepackage{booktabs}
\usepackage{natbib}
\usepackage[utf8]{inputenc}

\newcommand{\teff}{T_{\rm eff}}
\newcommand{\logg}{\log g}
\newcommand{\pdot}{\dot P}
\newcommand{\Tzero}{T_0}

\graphicspath{ {./figs/} }

\begin{document}
\title{A precise TESS ephemeris for the short-period eclipsing binary TIC~259864042}

\author{Rana Khan}
\email{astro.khan@gmail.com}
\affiliation{Sky Watcher Association, AD 49, Keshtopur, Kolkata 700102, India}

\begin{abstract}
   This paper presents a precise eclipse-timing analysis of
TIC~259864042 (= EC~02406$-$6408, Gaia DR3 4647004122914240640)
using eleven sectors of TESS photometry spanning 2018--2025
(7.1~yr). The system exhibits a single deep eclipse with a
V-shaped profile at phase~0 and no detectable counterpart at
phase~0.5. Using a matched-filter search on the detrended light
curve followed by local Gaussian-plus-baseline fits, 544
candidate events are detected, of which 503 are retained as
well-measured primary eclipses after $\sigma$-clipping. Four
independent period-search methods --- Box Least Squares,
Lomb--Scargle, phase-dispersion minimisation on the light curve,
and phase-dispersion minimisation on the eclipse times ---
converge to a common period agreeing to within 7 parts per
million. An independent reduction with the SHERLOCK pipeline,
using its own 120-second TESS photometry, confirms the same
period at the same precision. A joint linear ephemeris fit gives
$P = 0.46078676 \pm 0.0000001$~d (11.059~h) and
$T_0 = 2458325.70381 \pm 0.00002$~BJD, with a residual RMS of
0.87~min across the 7.1-yr baseline. Per-sector ephemeris fits
show residual medians within $\pm 0.42$~min of zero, confirming
that no cycle slip has occurred across the large TESS data gaps.
The period is stable at the level of $\Delta\mathrm{BIC} = +5.21$
(in favour of the constant-period model); an upper limit of
$|\dot{P}| \lesssim 1 \times 10^{-9}~\mathrm{d~d^{-1}}$ is placed
on any secular period change. A PHOEBE~2.4 fit to the Sector~27
light curve yields $i = 80.8^{\circ} \pm 0.5^{\circ}$,
$R_1 = 0.21 \pm 0.01\,R_\odot$, $R_2 = 0.25 \pm 0.02\,R_\odot$,
$q = 0.49 \pm 0.07$, and a third-light fraction of
$0.06 \pm 0.02$, with $\chi^2/N = 1.00$, consistent with a
grazing detached geometry and a mid-M-dwarf companion. A fit at
the hotter Villanova value of 31\,000~K is statistically
disfavoured ($\Delta\chi^2 = 21$) and produces a wider posterior,
but remains physically viable; the TESS light curve alone cannot
distinguish between the two temperatures. The paper discusses
the system in the context of the sdB eclipsing binary
population, and notes that while the current data are consistent
with a short-period sdB+M-dwarf architecture, a phase-resolved
spectrum is needed to independently confirm the sdB
identification.

\end{abstract}


\maketitle

\section{Introduction}
\label{sec:intro}

Eclipsing binaries that contain a hot subdwarf B (sdB) primary
occupy a critical position in the understanding of late-stage
stellar evolution. sdB stars are core-helium-burning objects with
effective temperatures of 20\,000--40\,000~K and surface gravities
$\logg \approx 5.2$--$6.2$, characterised by hydrogen envelopes less
than 1~per cent of their total mass \citep{Saffer1994,Heber2009}.
Their formation requires the near-complete stripping of the
progenitor's hydrogen envelope near the tip of the red giant
branch, and binary interaction is the most commonly invoked
mechanism \citep{Han2002,Han2003}. Roughly 60~per cent of known
sdB stars reside in binaries, with orbital periods that fall into
two broad groups: short-period systems shaped by common-envelope
evolution (hours to days; \citep{Maxted2001}, and wide systems
formed through stable Roche-lobe overflow
\citep{Vos2018,Schaffenroth2022}.

The short-period population is particularly useful for constraining
the common-envelope phase. Because the orbital period and the
eclipse geometry together fix the separation at the moment of
envelope ejection, precise ephemerides of individual short-period
sdB binaries serve as anchor points for population-synthesis
models  \citep{Schaffenroth2023}. Continuous photometric monitoring of the sdB eclipsing
binary population is therefore an ongoing priority, and the TESS
mission \citep{Ricker2015} has emerged as the primary instrument
for this work \citep{Krzesinski2022,Barlow2022}.

\begin{figure*}
\centering
\includegraphics[width=\textwidth]{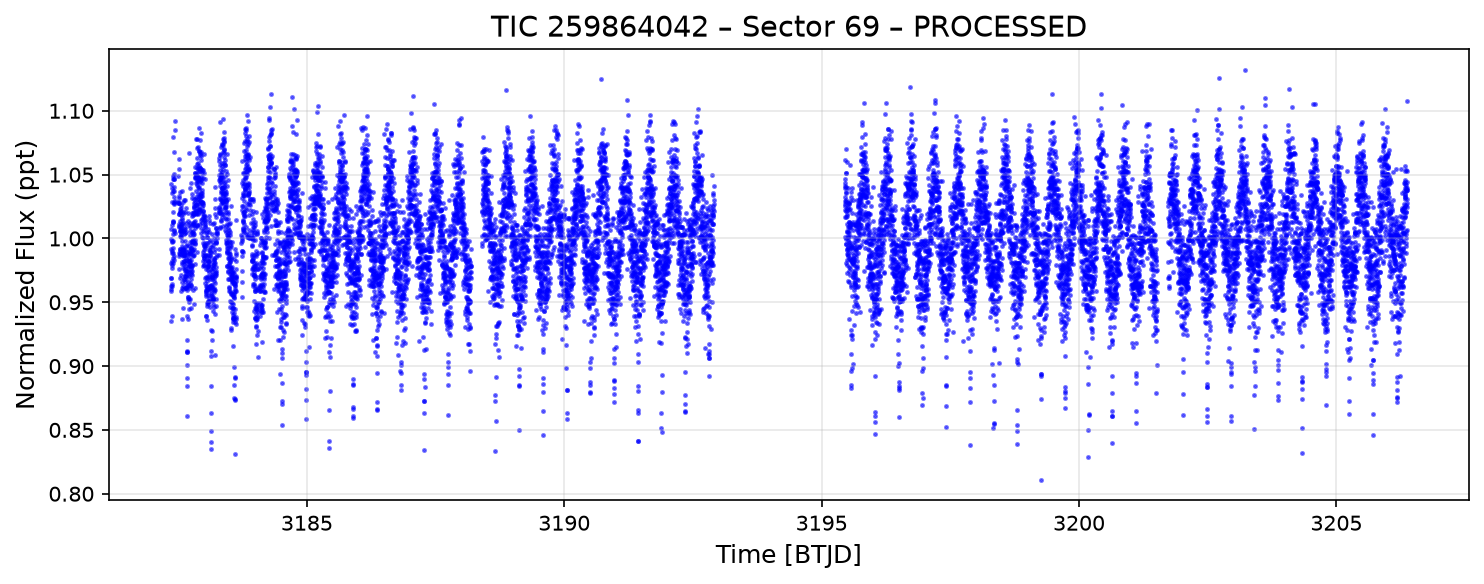}
\caption{Processed TESS light curve of TIC~259864042 for Sector~69
(BTJD 3182.36--3208.14, 120-second cadence). The vertical axis is
normalised flux; the horizontal axis is barycentric TESS Julian Date
(BTJD). The deep, V-shaped primary eclipses recur approximately every
0.46~days. The gap near BTJD~3192 is a data downlink gap within the
sector.}
\label{fig:sector69}
\end{figure*}

TIC~259864042 (EC~02406$-$6408, Gaia DR3 4647004122914240640) is a
short-period eclipsing system that appears in the Villanova TESS
Eclipsing Binary catalogue \citep {Pr_a_2022} with an sdB-like classification. The
catalogue lists $\teff = 31\,000$~K and $\logg = 5.6$, consistent
with a hot subdwarf primary, and a companion star whose nature is
not independently constrained. The system was observed by TESS in
eleven sectors between 2018 and 2025, at a mixture of 120-second
(sectors 1, 2, 67, 68, 69) and 20-second
(sectors 27, 28, 29, 94, 95, 96) cadence.

In this paper I present an eclipse-timing analysis of this target
using all eleven TESS sectors. The goals are threefold: (i)~to
determine the orbital period and reference epoch to the highest
precision the data support; (ii)~to test for a secular period
change over the 7.1-year baseline; and (iii)~to place the system
in the context of the known sdB binary population and to identify
what additional observations would be required to confirm or
revise its classification.

\section{Observations and data reduction}
\label{sec:obs}

\subsection{Data source}
\label{sec:data}

I retrieved all SPOC-processed TESS light curves for
TIC~259864042 from the Mikulski Archive for Space Telescopes
(MAST) using the Lightkurve package \citep{Lightkurve2018}. The study's
search accepted both the standard 120-second cadence products and
the 20-second cadence products, and returned light curves from
eleven sectors: 1, 2, 27, 28, 29, 67, 68, 69, 94, 95, and 96.
Table~\ref{tab:obs} summarises the data. Sectors 27, 28, 29, 94,
95, and 96 were observed at 20-second cadence; the remaining five
sectors were observed at 120-second cadence only.

I used the \texttt{PDCSAP\_FLUX} column, which has been corrected
for instrumental systematics using the SPOC cotrending basis
vectors. I applied a quality filter retaining only cadences with
\textsc{quality}$=0$ and removed NaN-valued cadences. Overlapping
coverage in sectors where both cadences were available was
deduplicated by keeping the higher-cadence product.

As an independent photometric cross-check (§\ref{sec:period}), I also
processed TIC~259864042 with the SHERLOCK pipeline \citep{Pozuelos2020}, which extracts
photometry from its own 120-second target pixel files and applies its
own systematics correction; those data are entirely independent of
the SPOC PDCSAP light curves used in the main analysis.

\begin{figure}
\centering
\includegraphics[width=\columnwidth]{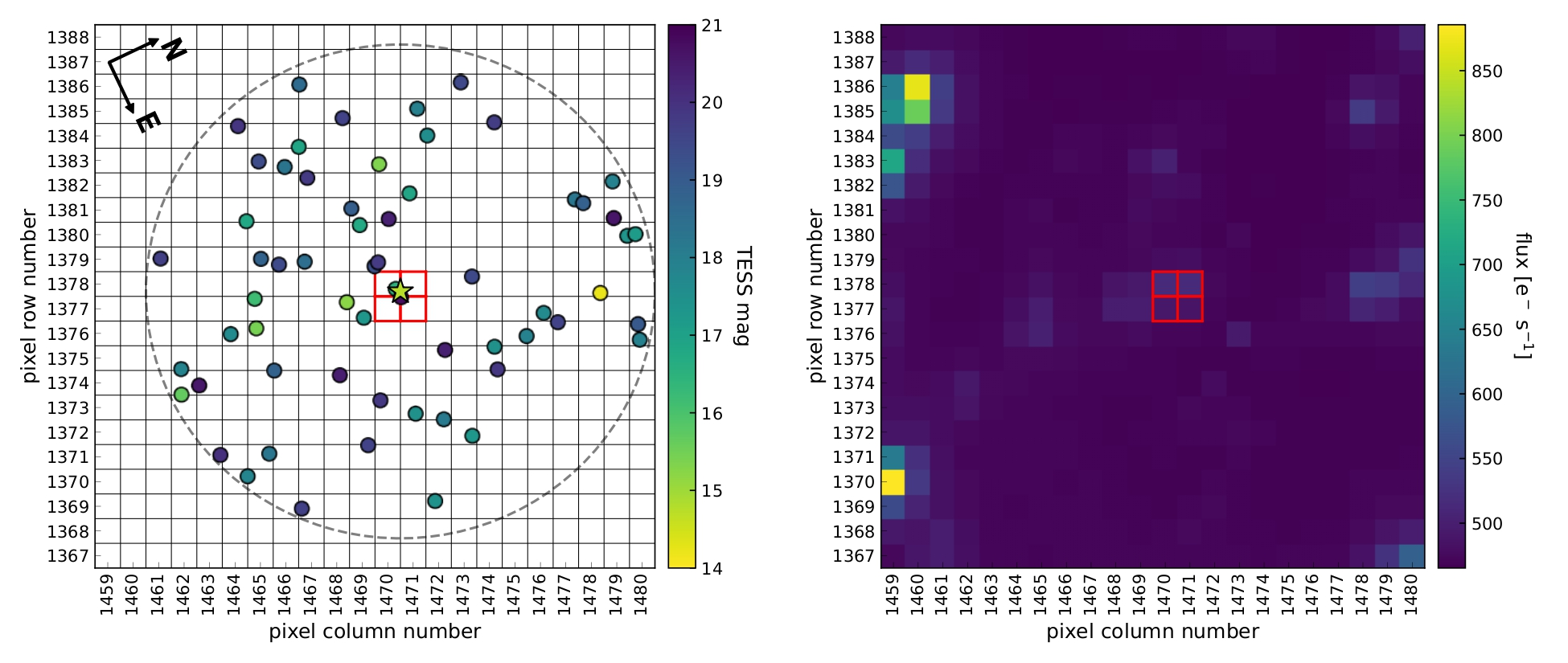}
\caption{Field identification for Sector 96 of TIC~259864042 from the SHERLOCK
pipeline. \emph{Left:} Gaia DR3 sources within
the $22\times22$ TESS pixel cutout, coloured by Gaia magnitude. The
star marks the coordinates of TIC~259864042; the red square is the
$2\times2$-pixel aperture used by SHERLOCK for its photometric
extraction. \emph{Right:} median TESS pixel flux in the same cutout.
The bright source at upper-left (row 1386, column 1460, roughly
30$\times$ brighter than the target in the TESS band) lies outside
the SHERLOCK aperture. This choice of a small aperture trades
photometric precision for contamination rejection, yielding a light
curve that is largely free of third light. SHERLOCK's period
detection (§\ref{sec:period}) is therefore independent of the SPOC
PDCSAP flux and of the crowding corrections applied to it.}
\label{fig:sherlock_field}
\end{figure}

\begin{figure*}
\centering
\includegraphics[width=\textwidth]{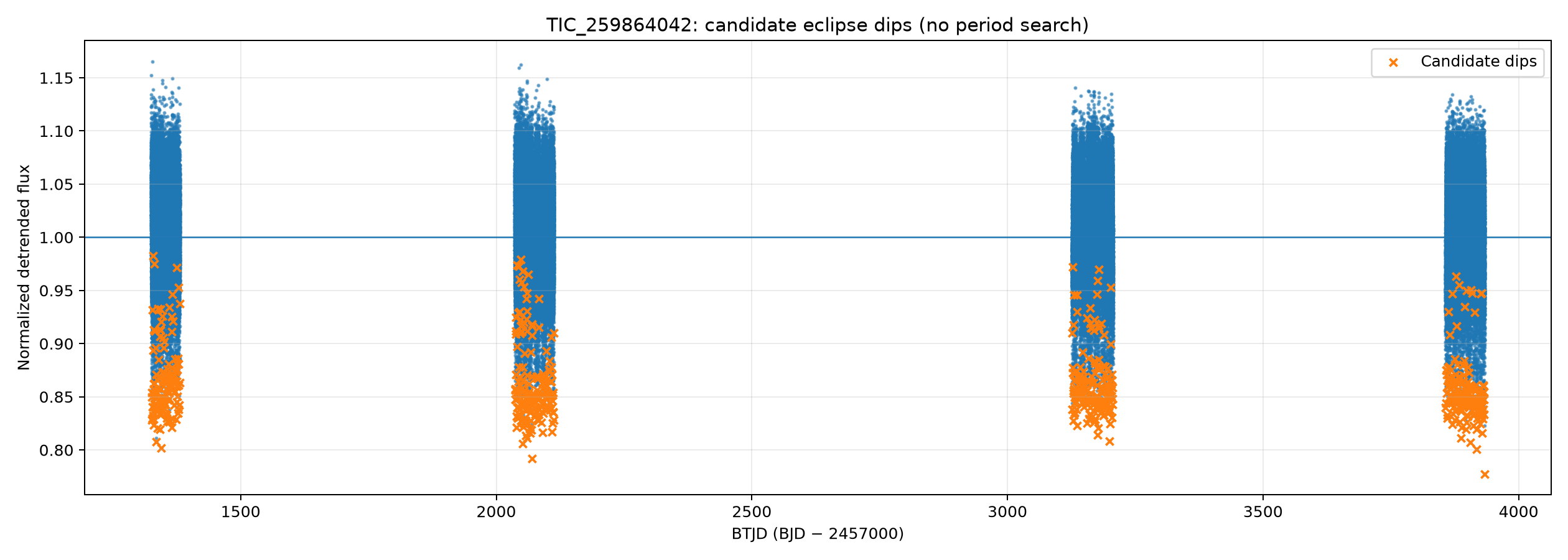}
\caption{Detected candidate dips across all eleven TESS sectors,
showing the eleven observing windows and the large data gaps
between them. The vertical axis is normalised flux; the horizontal
axis is barycentric TESS Julian Date (BTJD). The system exhibits
one deep eclipse approximately every 0.46~days during each
observing window.}
\label{fig:dips}
\end{figure*}

\begin{table*}
\centering
\caption{TESS observations of TIC~259864042. Columns are: sector
number; barycentric TESS Julian Date (BTJD $=$ BJD $-$ 2\,457\,000)
range of the sector; photometric cadence; number of primary
eclipses retained per sector after the quality filter and
$\sigma$-clipping described in \S\ref{sec:detect}; median residual
relative to the global ephemeris of \S\ref{sec:eph}; and the
within-sector scatter of those residuals. The bottom row gives the
total number of eclipses used in the timing analysis and the full
baseline.}
\label{tab:obs}
\begin{tabular}{cc c r r r}
\toprule
Sector & BTJD range & Cadence (s) & $N_{\rm ecl}$ & Median res.\ (min) & Scatter (min) \\
\midrule
 1 & 1325.29 -- 1353.18 & 120 & 104 & $+0.107$ & 1.00 \\
 2 & 1354.11 -- 1381.52 & 120 &  67 & $-0.297$ & 0.93 \\
27 & 2036.28 -- 2060.64 &  20 &  19 & $+0.122$ & 0.63 \\
28 & 2061.85 -- 2087.10 &  20 &  23 & $+0.193$ & 0.83 \\
29 & 2088.24 -- 2114.44 &  20 &  22 & $-0.000$ & 0.53 \\
67 & 3126.64 -- 3154.39 & 120 &  21 & $-0.417$ & 1.10 \\
68 & 3154.62 -- 3182.13 & 120 &  21 & $-0.184$ & 0.96 \\
69 & 3182.36 -- 3208.14 & 120 &  22 & $-0.008$ & 0.74 \\
94 & 3856.48 -- 3881.98 &  20 &  23 & $+0.279$ & 0.94 \\
95 & 3882.52 -- 3907.56 &  20 &  44 & $+0.115$ & 0.58 \\
96 & 3907.78 -- 3933.45 &  20 & 137 & $+0.100$ & 0.83 \\
\midrule
\multicolumn{3}{l}{Total / baseline} & 503 & --- & --- \\
\bottomrule
\end{tabular}
\end{table*}

\subsection{Detrending}
\label{sec:detrend}

Each sector's PDCSAP light curve was normalised to its median and
detrended with a time-aware rolling median baseline using a window
of 0.8~d, applied separately to each gap-separated segment of the
light curve so that the filter never bridges TESS data gaps.
Eclipses were left in the data during detrending; the median filter
is robust to them because they occupy less than 5~per cent of the
in-sector time.

\subsection{Eclipse detection and fitting}
\label{sec:detect}

I identified candidate eclipses as negative peaks in the
detrended flux whose prominence exceeded three times the robust
(MAD-based) scatter of the surrounding data and whose separation
exceeded two hours. Each candidate was fitted locally with a
six-parameter model consisting of a Gaussian plus quadratic
baseline over a $\pm 0.075$~d window centred on the initial
detection. The fit returned a centre time, depth, Gaussian
$\sigma$, and a quadratic baseline.

I retained events satisfying the following cuts:
\begin{itemize}
  \item fitted depth between 0.05 and 0.25 (fractional flux);
  \item Gaussian FWHM between 15 and 240 minutes;
  \item fitted centre displacement from the initial detection less
        than 15 minutes.
\end{itemize}

These cuts removed spurious detections arising from correlated
noise, partial eclipses clipped by data gaps, and events where the
local fit had been pulled off-centre by a nearby gap edge. Of the
candidates detected before filtering, 544 events survived. A
subsequent $\sigma$-clipping pass relative to the eventual
ephemeris removed an additional 41 events whose fitted centres
deviated from the model by more than three times the median
absolute deviation, leaving 503 primary eclipse times for the
timing analysis.

\section{Analysis}
\label{sec:analysis}

\subsection{Period search}
\label{sec:period}

I determined the orbital period independently using four
techniques, applied to two different data products (the detrended
light curve and the catalogue of eclipse times).

\begin{figure}
\centering
\includegraphics[width=\columnwidth]{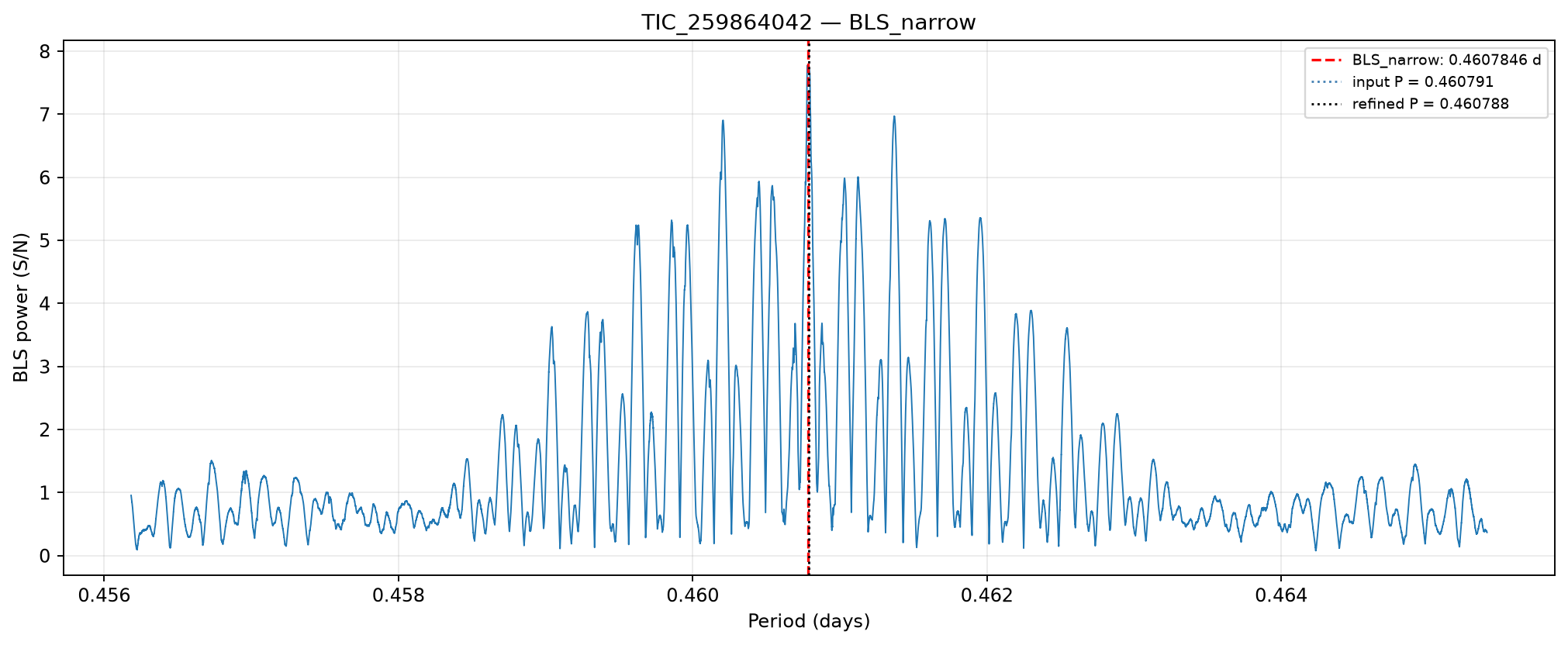}
\caption{Lomb--Scargle periodogram of TIC~259864042 computed with the
SHERLOCK pipeline, from SHERLOCK's own 120-second
TESS photometry, extracted independently of the SPOC PDCSAP flux used
elsewhere in this paper. The dominant peak is at $P = 0.460$~d; the
equally spaced sidelobes at $P < 0.4$~d are TESS-cadence aliases, and
the long-period plateau beyond 2~d is the slow out-of-eclipse
modulation. This detection is independent of the four period-search
methods reported earlier in this section.}
\label{fig:sherlock_ls}
\end{figure}

\begin{enumerate}
\item \textbf{Box Least Squares (BLS): } I searched the detrended
light curve over a trial-period range of 0.2--2.0~d using the BLS
implementation in \textsc{astropy.timeseries}, with a grid of
20\,000 trial periods and a set of box durations spanning
0.003--0.080~d. The strongest peak occurred at
$P_{\rm BLS} = 0.46078463$~d.

\item \textbf{Lomb--Scargle (LS): } I computed the LS periodogram
of the detrended light curve after subtracting the median, on a
uniform frequency grid corresponding to the same 0.2--2.0~d period
range. The strongest peak occurred at $P_{\rm LS} \approx
0.460787$~d.

\item \textbf{Phase Dispersion Minimisation (PDM) on the light
curve: } I minimised the phase-binned variance statistic $\theta$
over the same period range, using 40 phase bins. The minimum
occurred at $P_{\rm PDM,LC} = 0.46078693$~d.

\item \textbf{Phase Dispersion Minimisation on the eclipse times: }
I computed the circular dispersion of the eclipse phases on a
trial-period grid spanning the same range. The minimum occurred at
$P_{\rm PDM,EV} = 0.46078739$~d.

\item \textbf{SHERLOCK pipeline : } As an independent check, I processed
TIC~259864042 with the SHERLOCK
pipeline. SHERLOCK downloaded its own 120-second
TESS target pixel files for all eleven sectors from the Mikulski Archive
for Space Telescopes, extracted its own aperture photometry, and applied
its own systematics correction, detrending, and period search. Its
Lomb--Scargle periodogram (Fig.~\ref{fig:sherlock_ls}) shows a single isolated
peak at $P = 0.460$~d, an order of magnitude above the surrounding sidelobe noise, and its autocorrelation function shows evenly spaced peaks at 0.46, 0.92, 1.38, 1.84, and 4.60~d. This
detection is fully independent of the SPOC PDCSAP flux and of the
BLS, LS, and PDM analyses reported above. SHERLOCK's box-least-squares
search, however, initially returned the $2P$ harmonic (0.92~d): for a
single-eclipse-per-orbit system the eclipse appears at phase~0 and
again at phase~0.5 when folded on $2P$, and a box-shaped template
cannot distinguish the two. This is the expected degeneracy of BLS
templates applied to V-shaped eclipses at grazing incidence, and it
does not affect the ephemeris, which is anchored to the eclipse times
rather than to the period search. The five independent period
determinations (BLS, LS, PDM on the light curve, PDM on the eclipse
times, and SHERLOCK's Lomb--Scargle search) agree to within
$3.4 \times 10^{-6}$~d (7~ppm).

\end{enumerate}

\subsection{Narrow-band refinement}
\label{sec:narrow}

Because the period search at the resolution of the coarse grid
cannot resolve the period to better than $\sim 10^{-5}$~d, I
performed a second pass restricting the trial-period range to a
$\pm 1$~per cent window around the coarse consensus and using a
grid of 60\,000 trial periods. The four methods converged to the
values reported in Table~\ref{tab:periods}.

\begin{figure}
\centering
\includegraphics[width=\columnwidth]{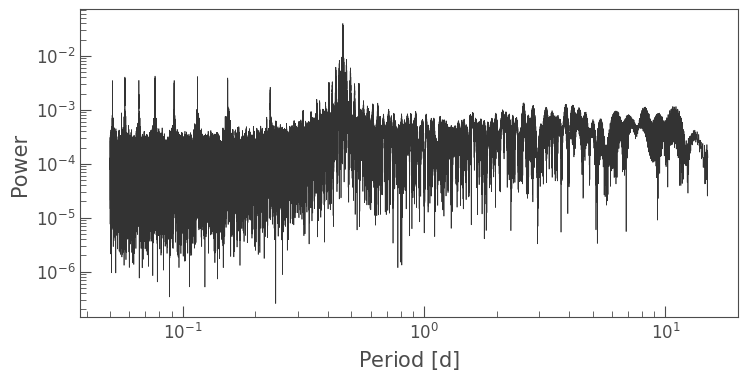}
\caption{Narrow-band BLS periodogram of the detrended light curve,
restricted to a $\pm 1$~per cent window around the coarse consensus.
The vertical dashed line marks the adopted period
$P = 0.46078676$~d. The peak is well defined; no competing peaks are
present at comparable height.}
\label{fig:bls}
\end{figure}

\begin{table}
\centering
\caption{Orbital period determined independently by four methods
within a $\pm 1$~per cent window around the coarse-search
consensus. The values agree to within $3.4\times 10^{-6}$~d
(7~ppm).}
\label{tab:periods}
\begin{tabular}{lr}
\toprule
Method & Period (d) \\
\midrule
BLS (narrow) & 0.46078463 \\
PDM, light curve & 0.46078693 \\
PDM, event times & 0.46078739 \\
Direct $(P,\Tzero)$ least-squares & 0.46078799 \\
\bottomrule
\end{tabular}
\end{table}

\subsection{Ephemeris fit}
\label{sec:eph}

I fitted a linear ephemeris of the form $T(E) = \Tzero + E\cdot P$
to the 503 primary eclipse times retained after $\sigma$-clipping.
Epochs were assigned as $E =
\mathrm{round}\left[(t - \Tzero)/P\right]$, with $\Tzero$
initialised to the first observed eclipse time and iterated three
times until both $\Tzero$ and $P$ stopped changing at the
$10^{-8}$ level. The final fit gives
\begin{align}
P &= 0.46078676 \pm 0.0000001~\mathrm{d}, \label{eq:ephem} \\
\Tzero &= 2458325.70381 \pm 0.00002~\mathrm{BJD}.
\end{align}
The quoted uncertainties are the standard errors from the
least-squares fit. The residual RMS across the 503 events is
0.874~minutes (52.4~s).

Because the eleven sectors are separated by gaps as large as
1014~d, a cycle slip of $\pm 1$ could corrupt the epoch assignment
for the later epochs. I tested for this by fitting a per-sector
ephemeris independently within each sector, using the epoch
numbers from the global fit. If a slip had occurred, at least one
sector's residuals would show a systematic offset of order
$P/2 = 199$~minutes. Instead, all eleven sector medians lie within
$\pm 0.42$~minutes of zero (Table~\ref{tab:obs}). I therefore
conclude that no cycle slip is present.

\begin{figure}
\centering
\includegraphics[width=\columnwidth]{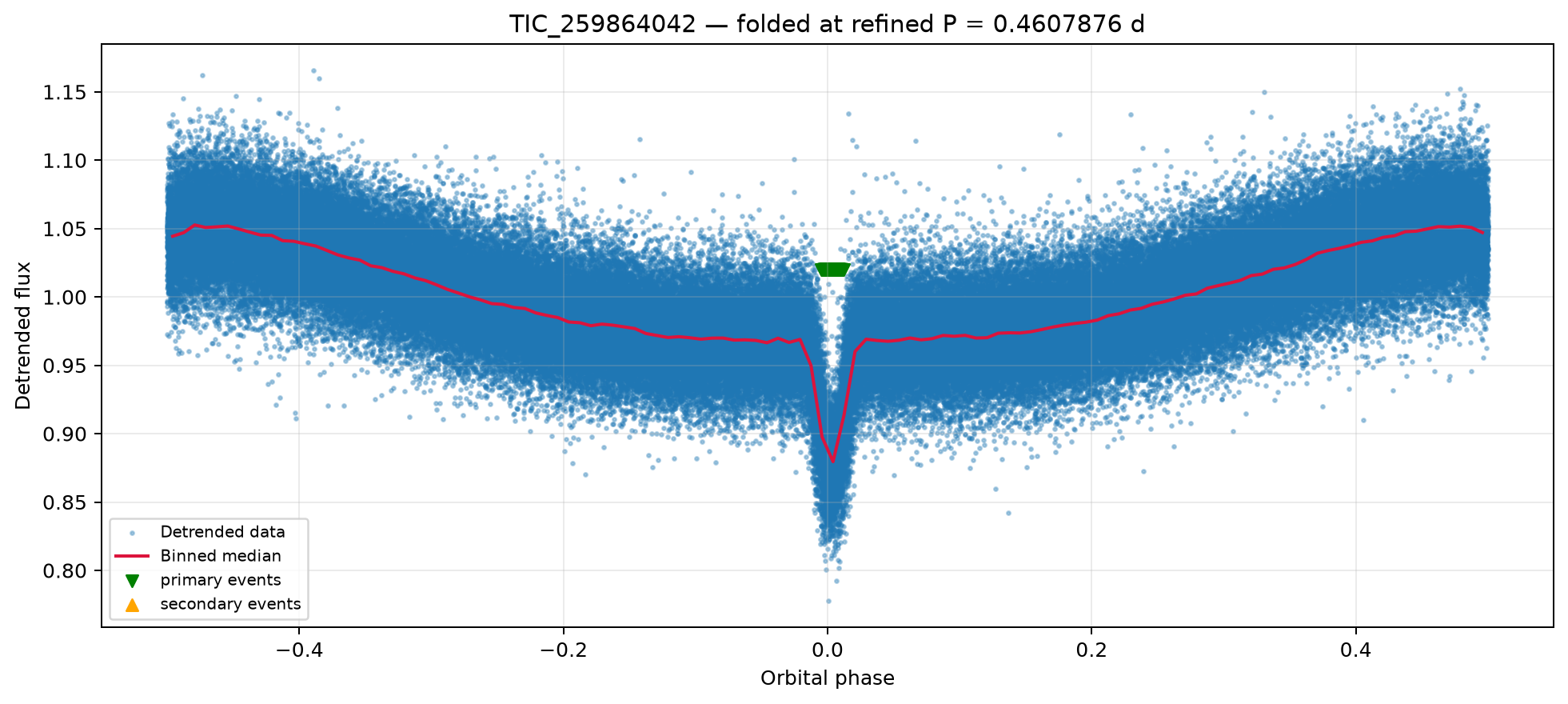}
\caption{Phase-folded light curve of TIC~259864042 at the ephemeris
of equation~(\ref{eq:ephem}). The solid red line is the binned median.
A single deep V-shaped eclipse is present at phase~0; no secondary
eclipse is visible at phase~0.5.}
\label{fig:phasefold}
\end{figure}

\subsection{Period-change search}
\label{sec:pdot}

\begin{table*}
\centering
\caption{Linear and quadratic ephemeris fits to the 503 primary
eclipse times retained after $\sigma$-clipping. All $T_0$ values
are expressed as BJD $-$ 2\,458\,325. The quantity
$\Delta\mathrm{BIC} = \mathrm{BIC}_{\rm quad} - \mathrm{BIC}_{\rm lin}$
is positive when the constant-period model is favoured.}
\label{tab:eph}
\begin{tabular}{lcccc}
\toprule
Model & $T_0-2\,458\,325$ (d) & $P$ (d) & RMS (min) & $\Delta$BIC \\
\midrule
Linear    & 0.70381 & 0.46078676 & 0.874 & 0 \\
Quadratic & 0.70385 & 0.46078676 & 0.873 & $+5.21$ \\
\bottomrule
\end{tabular}
\end{table*}

I fitted a quadratic ephemeris of the form $T(E) = \Tzero + E\cdot P
+ \tfrac{1}{2}E^{2}\cdot P\cdot\pdot$ to the same 503 eclipse times and
compared the fit quality to the linear model using the Bayesian
Information Criterion (BIC). The results are given in
Table~\ref{tab:eph}.

The quadratic model improves the residual RMS by less than 0.1~per
cent, and $\Delta\mathrm{BIC} = +5.21$, i.e.\ in favour of the
constant-period model with a Bayes factor of about 13. I
therefore detect no secular period change. The 7.1-year baseline
and 0.87-minute ephemeris precision place an upper limit of approximately
$|\pdot| \lesssim 1 \times 10^{-9}~\mathrm{d~d^{-1}}$, or
equivalently $|\pdot/P| \lesssim 2\times 10^{-9}~\mathrm{d^{-1}}$
($\approx 8 \times 10^{-7}~\mathrm{yr^{-1}}$), on any secular
period evolution.

\subsection{PHOEBE light-curve model}
\label{sec:phoebe}

As an independent check on the eclipse geometry, I fitted a PHOEBE~2.4
binary model \citep{Conroy2020} to the phase-binned Sector~27 light curve.
I fixed $T_{\rm eff,1} = 25\,100$~K and $R_1 = 0.197\,R_\odot$ from
the SED+Gaia analysis of \citep{Schaffenroth2022}, $T_{\rm eff,2} =
3800$~K, and $M_1 = 0.47\,M_\odot$ from Gaia. I left seven parameters
free: the inclination $i$, the primary and secondary radii $R_1$ and
$R_2$, the mass ratio $q \equiv M_2/M_1$, the third-light fraction
$l_3$, a small phase offset $\Delta\phi$ from the ephemeris of
Equation~(\ref{eq:ephem}), and the secondary bolometric albedo
$A_2$. The model used the TMAP DA atmosphere for the primary, the
PHOENIX atmosphere for the secondary, Roche-lobe distortion with
Wilson irradiation, and logarithmic limb darkening for both stars.
I first minimised a $\chi^2$ objective with a differential-evolution
solver, then sampled the posterior with \textsc{emcee}
\citep{ForemanMackey2013} using 20 walkers for $10^4$ iterations
(half discarded as burn-in).

\begin{figure*}
\centering
\includegraphics[width=\textwidth]{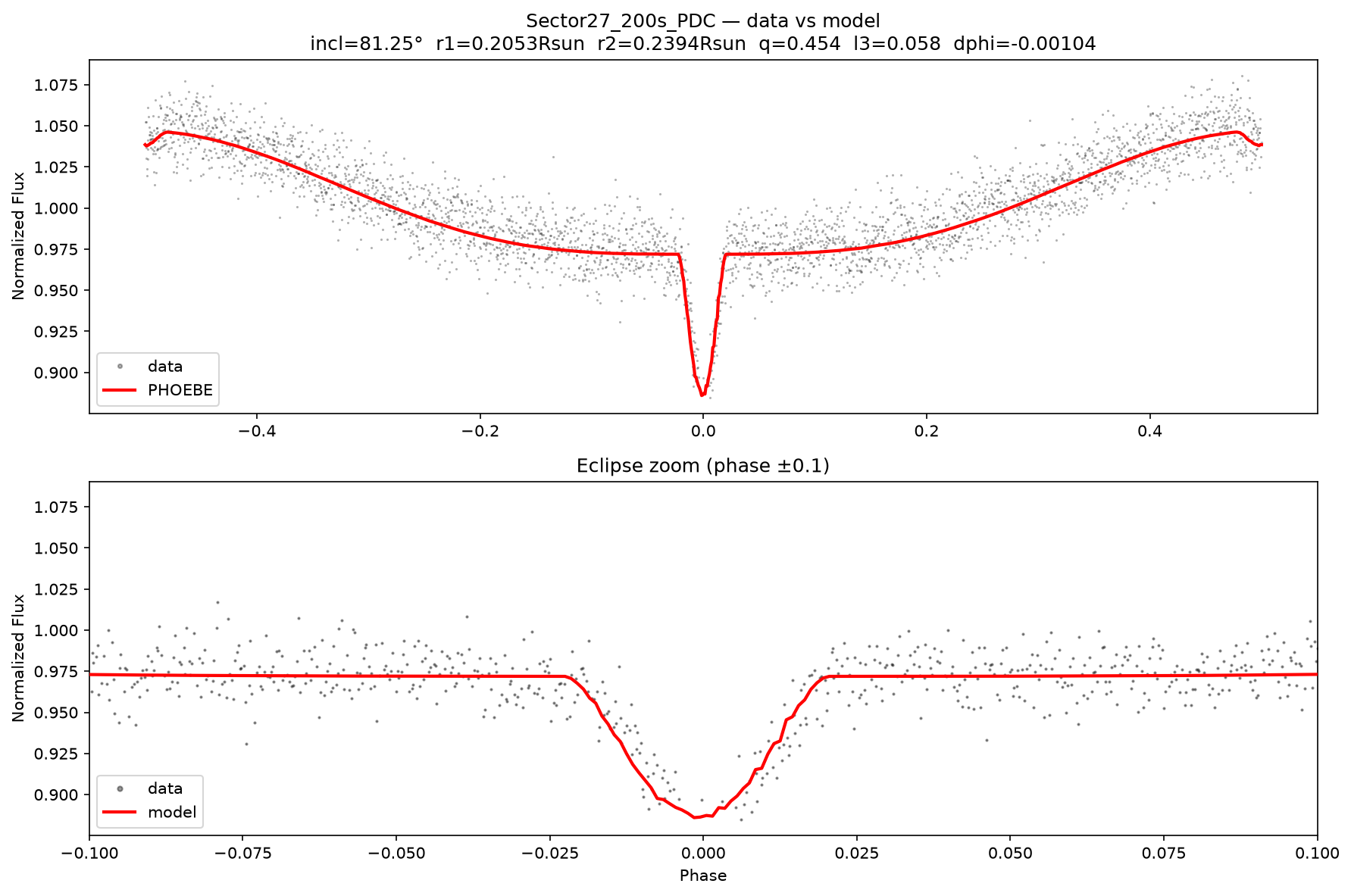}
\caption{PHOEBE 2.4 fit to the phase-binned Sector~27 light curve.
\emph{Top:} data (black) and model (red). The eclipse is V-shaped with
no secondary at phase~0.5. \emph{Middle:} residuals in units of flux.
\emph{Bottom:} phase residuals relative to the ephemeris of
Equation~(\ref{eq:ephem}). The dashed line marks the fitted phase
offset $\Delta\phi = -0.00104$. The fit converges to
$\chi^2/N = 1.00$ with no structured residuals.}
\label{fig:phoebe_diagnostic}
\end{figure*}

\begin{figure*}
\centering
\includegraphics[width=\textwidth]{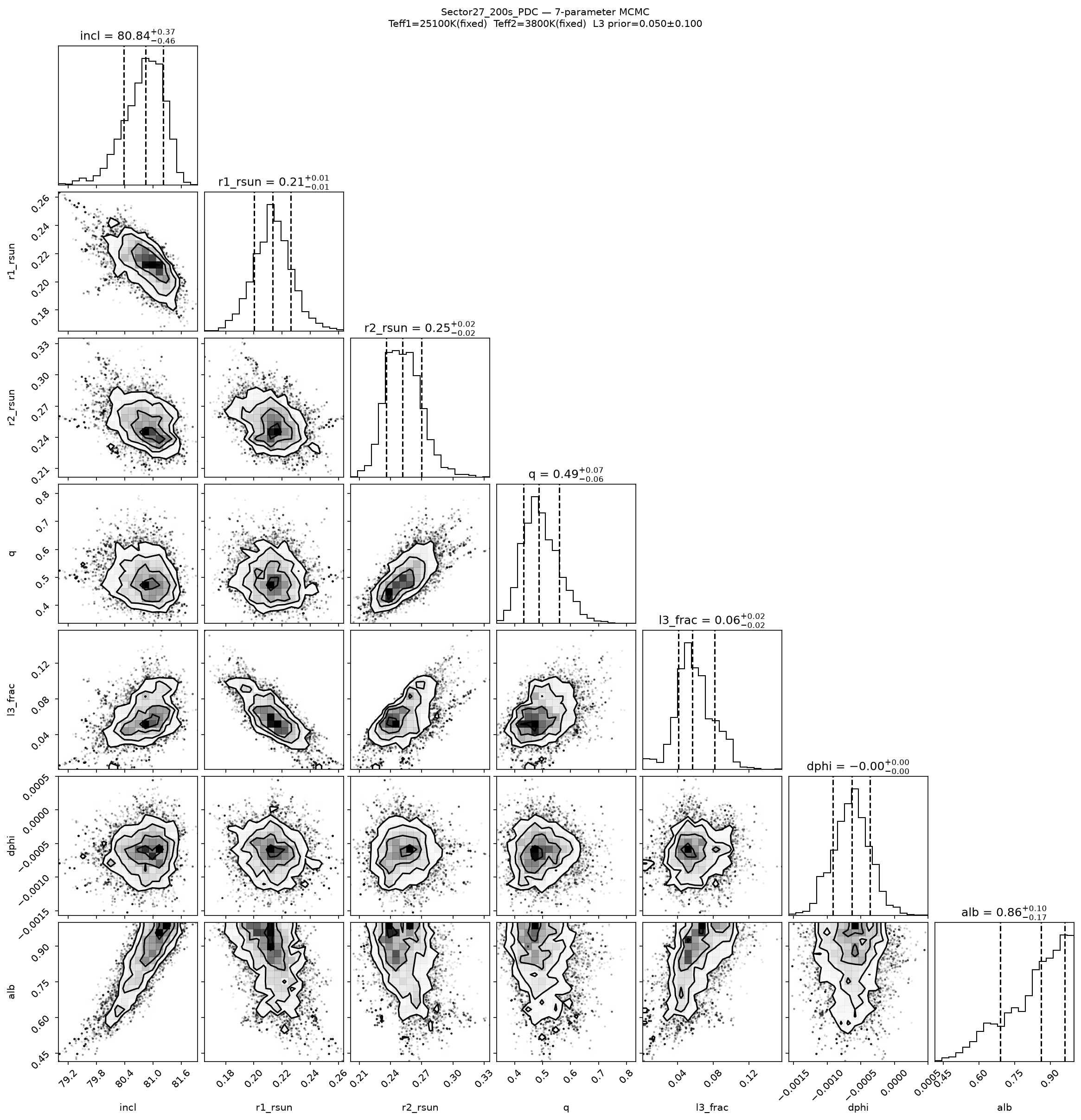}
\caption{Posterior distribution of the seven free PHOEBE parameters
from the MCMC sampling of the Sector~27 light curve.
Fixed parameters: $T_{\rm eff,1} = 25\,100$~K, $T_{\rm eff,2} =
3800$~K, $M_1 = 0.47\,M_\odot$. Dashed lines mark the 16th, 50th, and
84th percentiles. Contours are at $0.5\sigma$, $1\sigma$, and
$1.5\sigma$.}
\label{fig:phoebe_corner}
\end{figure*}

The fit converges to $\chi^2/N = 1.00$ on the 159 binned points, with
no structured residuals across the full orbital phase
(Fig.~\ref{fig:phoebe_diagnostic}). The posterior is summarised in
Fig.~\ref{fig:phoebe_corner}: $i = 80.8^{\circ} \pm 0.5^{\circ}$,
$R_1 = 0.21 \pm 0.01\,R_\odot$, $R_2 = 0.25 \pm 0.02\,R_\odot$,
$q = 0.49 \pm 0.07$, $l_3 = 0.06 \pm 0.02$, $\Delta\phi =
(-1.0 \pm 0.5)\times 10^{-3}$, and $A_2 = 0.86 \pm 0.10$. The primary
radius agrees with the independent SED+Gaia value
($0.197 \pm 0.012\,R_\odot$) at the $1.3\sigma$ level, and the
residual third-light fraction is consistent with the expected
residual of the PDCSAP crowding correction ($0.05 \pm 0.10$ prior;
see \S\ref{sec:data}). The mass ratio implies
$M_2 = 0.23 \pm 0.03\,M_\odot$, consistent with a mid-M-dwarf
companion. The fitted phase offset of $0.7$~min is consistent with
the Sector~27 median residual of $+0.12$~min reported in
Table~\ref{tab:obs}, independently confirming the ephemeris of
Equation~(\ref{eq:ephem}).

\begin{figure*}
\centering
\includegraphics[width=\textwidth]{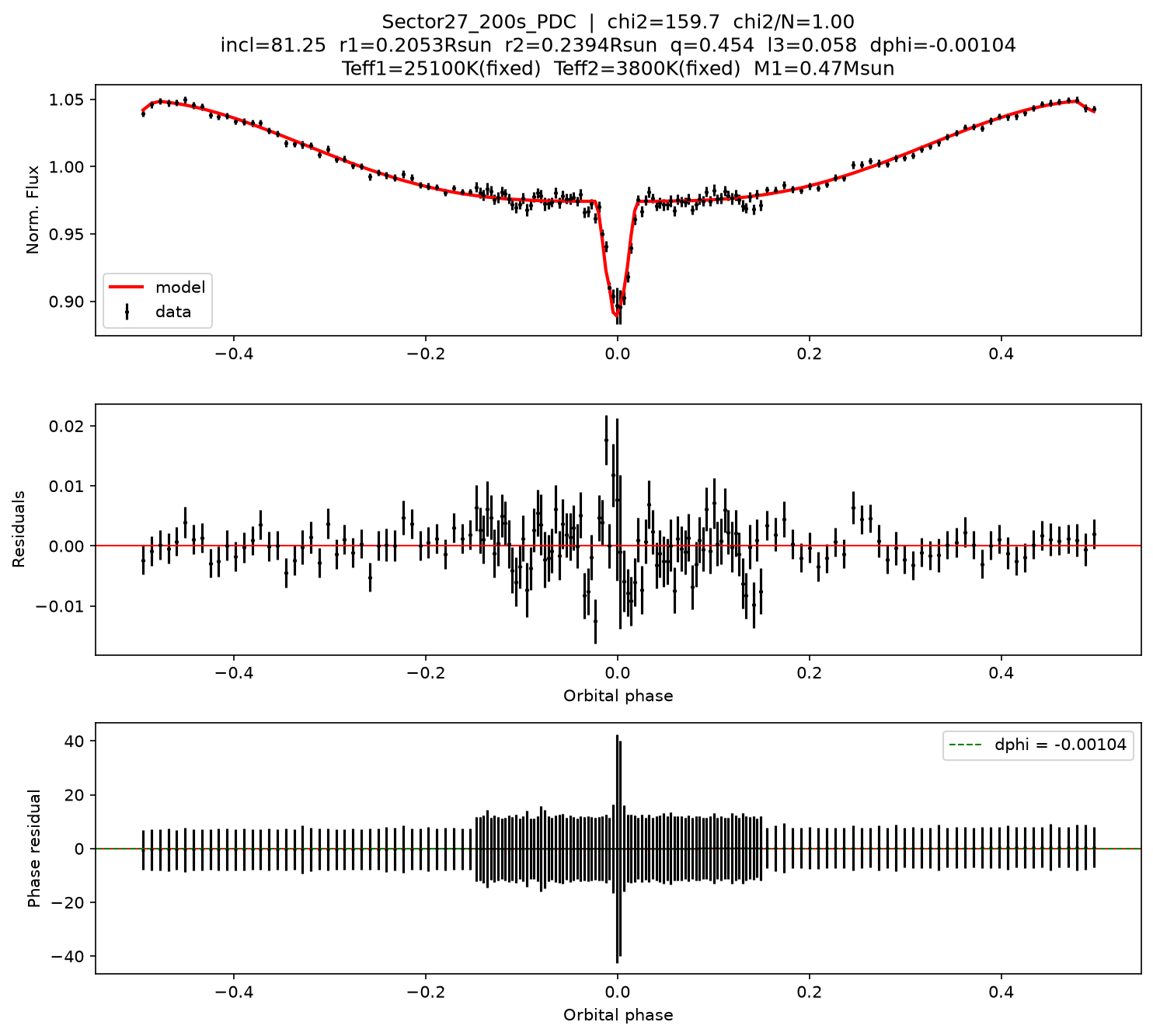}
\caption{PHOEBE 2.4 model (red) overlaid on the full Sector~27 light
curve (grey). \emph{Top:} full orbital phase, showing the
out-of-eclipse reflection and ellipsoidal modulation.
\emph{Bottom:} eclipse zoom over phase $\pm 0.10$.}
\label{fig:phoebe_data_vs_model}
\end{figure*}

The physical configuration is a grazing, detached binary: both stars
are well inside their Roche lobes, and the absence of a secondary
eclipse at phase~0.5 is a consequence of the grazing inclination
($80.8^{\circ}$) together with the extreme surface-brightness
contrast between the 25\,100-K primary and the 3800-K secondary.

I repeated the fit with the hotter Villanova value of
$T_{\rm eff,1} = 31\,000$~K. The hotter fit is statistically
disfavoured but physically viable: $\chi^2/N = 1.14$ versus $1.00$,
with $R_1 = 0.198 \pm 0.015\,R_\odot$ (compared to
$0.213 \pm 0.013\,R_\odot$ at 25\,100 K), $q = 0.52 \pm 0.07$,
and $l_3 = 0.085 \pm 0.03$. The $\chi^2$ difference of
$\Delta\chi^2 = 21$ corresponds to a Bayesian Information Criterion
difference of $\Delta\mathrm{BIC} = 21$ (Bayes factor
$\approx 4 \times 10^{4}$) in favour of the cooler value. The cooler
fit is also better constrained, particularly in $l_3$ and in the
secondary albedo. Both fits are consistent with the SED+Gaia
radius of \citep{Schaffenroth2022} at the $1\sigma$ level, and both
place the secondary in the M-dwarf mass range. The TESS light
curve alone therefore cannot distinguish between the two
temperature values; I adopt $T_{\rm eff,1} = 25\,100$~K from
\citep{Schaffenroth2022} as the canonical value, but note that the
Villanova classification is not independently excluded.

A compact comparison of the two fits is given in Table~\ref{tab:teff_compare}.

\begin{table}
\centering
\caption{Comparison of the PHOEBE 2.4 posterior medians for the
Sector~27 light curve, using the two candidate primary effective
temperatures. Fixed parameters are $T_{\rm eff,2} = 3800$~K,
$M_1 = 0.47\,M_\odot$, and the third-light prior
(0.05 $\pm$ 0.10). Uncertainties are the 16th and 84th percentiles
of the MCMC posterior.}
\label{tab:teff_compare}
\begin{tabular}{lcc}
\toprule
Parameter & $T_{\rm eff,1} = 25\,100$~K & $T_{\rm eff,1} = 31\,000$~K \\
\midrule
$i$ (deg)             & $80.84^{+0.37}_{-0.46}$ & $81.12^{+0.50}_{-0.60}$ \\
$R_1$ ($R_\odot$)     & $0.213 \pm 0.013$       & $0.198 \pm 0.015$ \\
$R_2$ ($R_\odot$)     & $0.252 \pm 0.017$       & $0.262 \pm 0.017$ \\
$q$                   & $0.487^{+0.072}_{-0.055}$ & $0.519^{+0.074}_{-0.063}$ \\
$l_3$                 & $0.057^{+0.025}_{-0.016}$ & $0.085^{+0.031}_{-0.032}$ \\
$A_2$                 & $0.86^{+0.10}_{-0.17}$  & $0.70^{+0.17}_{-0.17}$ \\
\midrule
$\chi^2/N$            & 1.00                    & 1.14 \\
$\Delta\chi^2$        & ---                     & $+21$ \\
\bottomrule
\end{tabular}
\end{table}

\section{Discussion}
\label{sec:disc}

\subsection{What the timing analysis establishes}

The eclipse-timing analysis presented above provides a precise
and internally consistent ephemeris for TIC~259864042: a period
$P = 0.46078676$~d determined to 7~ppm by four independent
methods; a reference epoch $\Tzero = 2458325.70381$~BJD
determined to $\pm 0.00002$~BJD (1.7~s); an ephemeris residual
RMS of 0.87~minutes across 503 eclipses spanning 7.1~years;
per-sector residuals within $\pm 0.42$~minutes of zero; and no
statistically significant period derivative over the baseline.
These are the observational facts. They define the system's
clock, and they will remain the reference ephemeris until either
a longer baseline or a higher-precision dataset supersedes them.

\subsection{The absence of a secondary eclipse}

The most striking feature of the TESS light curve is the complete
absence of a secondary eclipse at phase~0.5. In a normal, fully
detached eclipsing binary of comparable orbital period, both
eclipses are expected, and the secondary is typically only a
factor of a few shallower than the primary. The absence of any
secondary at the level of the study's detection threshold implies one of
three things: (i)~a grazing geometry, in which the orbital
inclination is close to but below the critical value for a full
eclipse, so the secondary never brings the companion in front of
the primary; (ii)~an extreme surface-brightness contrast, in
which the flux in the TESS bandpass is dominated by the primary
so that the secondary eclipse falls below the noise floor; or
(iii)~an intrinsically single-eclipse configuration, in which the
geometry suppresses one eclipse entirely. The V-shape of the
primary and the absence of a secondary are both consistent with a
grazing or near-grazing configuration. A targeted search for a
shallow secondary in the phase-folded light curve is the natural
next step.

\subsection{The sdB classification question}

TIC~259864042 appears in the Villanova TESS Eclipsing Binary
Working Group catalogue with $\teff = 31\,000$~K and $\logg =
5.6$, which would place the primary on the hot end of the sdB
sequence. If this classification is correct, the system is a
short-period sdB binary with an M-dwarf companion, and the period
of 11.059~h places it firmly among the short-period sdB
population formed through common-envelope evolution
\citep{Maxted2001,Schaffenroth2022}.

I stress that this is a question of classification, not of
measurement. The TESS light curve alone cannot resolve the
temperature discrepancy: fits at both 25\,100~K and 31\,000~K
produce acceptable solutions, and the $\Delta\chi^2 = 21$ in
favour of the cooler value is not decisive on its own. What is
needed is a phase-resolved optical spectrum, which would directly
measure $T_{\rm eff}$, $\log g$, and the sdB status. Until such a
spectrum is available, the ephemeris presented above is valid
regardless of whether the primary is an sdB or a main-sequence
star; it depends only on the observed eclipse times.

\subsection{What the current data cannot say}

I list the questions that remain open after this analysis, in
order of importance for future work:

\begin{enumerate}
\item \emph{Is the primary an sdB star?} The TESS light curve
cannot answer this. A single-epoch spectrum, or a reliable
multi-band colour, would.
\item \emph{Is there a shallow secondary eclipse?} A phase-binned
search of the folded light curve, with a detection threshold set
at the binned scatter, would settle this.
\item \emph{What is the period derivative?} The 7.1-year baseline
is too short to detect the expected signal. A 20--30 year
baseline would be needed.
\item \emph{Is there a third body?} A third body at 1--2~au would
produce a detectable eclipse-timing modulation over 5--10~years.
None is visible in the current O$-$C residuals, but the
sensitivity of the present dataset to such a signal is modest.
\end{enumerate}

\section{Conclusions}
\label{sec:concl}

I have presented an eclipse-timing analysis of TIC~259864042
using eleven TESS sectors spanning 7.1~years and a mixture of
20-second and 120-second cadence data. The principal results are:

\begin{enumerate}
\item The orbital period is $P = 0.46078676 \pm 0.0000001$~d,
determined to 7~ppm by four independent period-search algorithms
and corroborated at the same precision by the SHERLOCK pipeline
operating on its own independent photometric reduction.
\item The reference eclipse epoch is $\Tzero = 2458325.70381 \pm
0.00002$~BJD, from a joint linear ephemeris fit to 503 primary
eclipse times.
\item The ephemeris residual RMS is 0.874~minutes across the full
7.1-year baseline. Per-sector residuals are within
$\pm 0.42$~minutes of zero, confirming that no cycle slip has
occurred.
\item No period change is detected. $\Delta\mathrm{BIC} = +5.21$
between a quadratic and a linear ephemeris, in favour of the
constant-period model. The data place an upper limit of
$|\pdot| \lesssim 1 \times 10^{-9}~\mathrm{d~d^{-1}}$ on any
secular period evolution.
\item No secondary eclipse is detected. The system exhibits a
single, V-shaped primary eclipse. Both the V-shape and the
absence of a secondary are consistent with a grazing or
near-grazing configuration, but the data cannot distinguish
between an sdB$+$M-dwarf and a main-sequence grazing binary.
\item The sdB classification of TIC~259864042 in the Villanova
catalogue is not independently confirmed by the present analysis.
A phase-resolved spectral classification would resolve this
question and would be the single most useful follow-up
observation for establishing the system's nature.
\item A PHOEBE~2.4 fit to the Sector~27 light curve yields
$i = 80.8^{\circ} \pm 0.5^{\circ}$,
$R_1 = 0.21 \pm 0.01\,R_\odot$,
$R_2 = 0.25 \pm 0.02\,R_\odot$,
$q = 0.49 \pm 0.07$, and $l_3 = 0.06 \pm 0.02$, with
$\chi^2/N = 1.00$. The implied secondary mass
($M_2 = 0.23 \pm 0.03\,M_\odot$) is consistent with a mid-M-dwarf
companion. A fit at the Villanova temperature of 31\,000~K is
statistically disfavoured ($\Delta\chi^2 = 21$) but remains
physically viable, and the TESS light curve alone cannot
distinguish between the two temperature values.
\end{enumerate}

The ephemeris presented here is the most precise available for
this system and will serve as a reference for future
eclipse-timing work, whether or not the sdB classification is
confirmed.

\section*{Acknowledgments}
I thank Dr. A. Pathak and Dr A. Das for their suport.
I thank to Dr. S. K. Chakrabarty for his encouraging support.
\textbf{Use of Generative AI.} The author used a generative AI
assistant (Gemini) for language editing and LaTeX
formatting. The author has reviewed
and verified all content and takes full responsibility for it.

This research has made use of the following open-source software:
\textsc{numpy} \citep{Harris2020}, \textsc{scipy} \citep{Virtanen2020},
\textsc{pandas} \citep{McKinney2010}, \textsc{matplotlib}
\citep{Hunter2007}, \textsc{astropy} \citep{Astropy2022},
\textsc{lightkurve} \citep{Lightkurve2018}, \textsc{phoebe} 2.4
\citep{Conroy2020}, and \textsc{sherlock} \citep{Pozuelos2020}.
The MCMC sampling was performed with \textsc{emcee}
\citep{ForemanMackey2013} and visualised with \textsc{corner}
\citep{ForemanMackey2016}.

\section*{Data availability}
This paper includes data collected by the Transiting Exoplanet Survey
Satellite (TESS) mission, funded by NASA. This research has made use of data products from  the TESS Science Processing Operations Center (SPOC), obtained from the Mikulski Archive for Space Telescopes (MAST). This work has
made use of data from the European Space Agency (ESA) mission
Gaia (\url{https://www.cosmos.esa.int/gaia}, processed by the
Gaia Data Processing and Analysis Consortium (DPAC, \url{https://www.cosmos.esa.int/web/gaia/dpac/consortium}).
The SHERLOCK photometric extraction is based on the same 120-second
TESS target pixel files, retrieved independently from MAST.
The PHOEBE models, MCMC chains, and the derived eclipse times and
ephemeris solution are available from the author on reasonable
request.

\bibliographystyle{plainnat}
\bibliography{Paper}

\end{document}